\documentclass[prl, preprintnumbers, twocolumn]{revtex4}
\usepackage{amsmath}
\usepackage{graphicx}
\usepackage{amssymb}

\newcommand{\be}{\begin{equation}}
\newcommand{\ee}{\end{equation}}
\newcommand{\bea}{\begin{eqnarray}}
\newcommand{\eea}{\end{eqnarray}}
\newcommand{\zmin}{z_{\rm min}}
\newcommand{\zmax}{z_{\rm max}}
\newcommand{\met}{\not{\!\!\rm E}_T}

\begin{document}

\preprint{ANL-HEP-PR-12-46, IIT-CAPP-12-08, MSUHEP-120702}

\title{Measuring Top Quark Polarization in Top Pair plus Missing Energy Events}

\author{Edmond L. Berger$^{b}$, Qing-Hong Cao$^{a}$,
Jiang-Hao Yu$^{c}$, Hao Zhang$^{b,d}$}
\affiliation{
\mbox{$^a$Department of Physics and State Key Laboratory of Nuclear Physics
and Technology, Peking University, Beijing 100871, China}  \\ 
\mbox{$^b$High Energy Physics Division, Argonne National Laboratory, Argonne, IL 60439, USA}  \\ 
\mbox{$^c$Department of Physics and Astronomy, Michigan State University, East Lansing, MI 48824, USA}\\
\mbox{$^d$Illinois Institute of Technology, Chicago, IL 60616-3793, USA}}

\begin{abstract}
The polarization of a top quark can be sensitive to new physics beyond the standard model. 
Since the charged lepton from top quark decay is maximally correlated with the top quark 
spin, it is common to measure the polarization from the distribution in the angle 
between the charged lepton and the top quark directions.   We propose a novel method 
based on the charged lepton energy fraction and  illustrate the method with a detailed simulation 
of top quark pairs produced in supersymmetric top squark pair production.  We show that the 
lepton energy ratio distribution that we define is very sensitive to the top quark polarization but 
insensitive to the precise measurement of the top quark energy.  
\end{abstract}

\maketitle

\noindent{\bf Introduction:~}Events with a top quark pair plus missing energy ($t\bar{t}+\!\met$) are promising channels in which to investigate models of new physics (NP) beyond the standard model (SM).  Missing energy originates typically from non-interacting or otherwise invisible dark matter (DM) candidates in the NP models, along with neutrinos from SM decays.  In these events the polarization of the top quark is sensitive to the chirality structure of the top quark's interaction with a postulated parent new heavy resonance and the DM.  The top quark polarization might provide a new way to gain insight into NP models.  Measurements of the top quark polarization tend to rely on the predicted angular correlation of the momentum of a charged lepton (from the top quark decay) with the top quark spin~\cite{Kane:1991bg}.  However, this measurement is difficult in $t\bar{t}+\met$ events because it is generally not possible to reconstruct the top quark kinematics, i.e., to  disentangle the kinematic effects of the DM particles from neutrinos that accompany the charged leptons in the top quark leptonic decay.  

In this Letter we define and examine the energy fraction of the charged lepton from the top quark as a novel measure of top quark polarization,  without the requirement of top quark reconstruction and knowledge of the dark matter mass and spin.  
We emphasize a few advantages of our energy ratio variable:  (i) it is sensitive to the top-quark polarization; (ii) it not sensitive to the mass splitting  between a heavy resonance parent and the DM candidate, provided that this splitting is not too small; (iii) the difference between the left-handed top-quark ($t_L$) and the right-handed ($t_R$) is not sensitive to the spin of a heavy parent resonance or to the collider energy. 

We illustrate our method with top squark ($\tilde{t}$) pair production in the minimal supersymmetric extension of the SM (MSSM), 
$pp \to \tilde{t} \overline{\tilde{t}} X \to t \bar{t} \tilde{\chi} \tilde{\chi} X$, where $\tilde{\chi}$ denotes a neutralino (the DM candidate) .     
Once a signal for a $\tilde{t}$
or another such NP candidate has been established,
an important next step would be to determine the characteristics of its interaction with the SM particles.   We do a full model simulation of 
$pp \to \tilde{t} \overline{\tilde{t}} X \to t \bar{t} \tilde{\chi} \tilde{\chi} X$ at the Large Hadron Collider energy 8 TeV, including typical experimental selection cuts~\cite{Cao:2012rz}.  From these events, we compute the energy fraction of the charged lepton from the top quark decays, 
show explicitly the relationship of this energy fraction to the top quark spin, and demonstrate what one may conclude about the nature of the 
interaction of the top quark and top squark from such data.   In the MSSM, the $t$-polarization probes the $\tilde{t}$-$t$-$\tilde{\chi}$ interaction and in turn the top squark mixing~\cite{Perelstein:2008zt}.

\noindent{\bf The method:~} In the leptonic decay of a top quark, $t\to b W^+ \to b \ell^+ \nu$, the 
correlation of the momentum of the charged lepton $\ell^+$ with the polarization $\hat{s}_t$ 
of the top quark, viewed in the top quark rest frame, takes the form 
$(1+\hat{s}_t z)/2$, where $z \equiv \cos\theta$ is the cosine of the angle between 
the top quark spin axis and the lepton momentum.   
\begin{figure}[t]
\includegraphics[scale=0.4,clip]{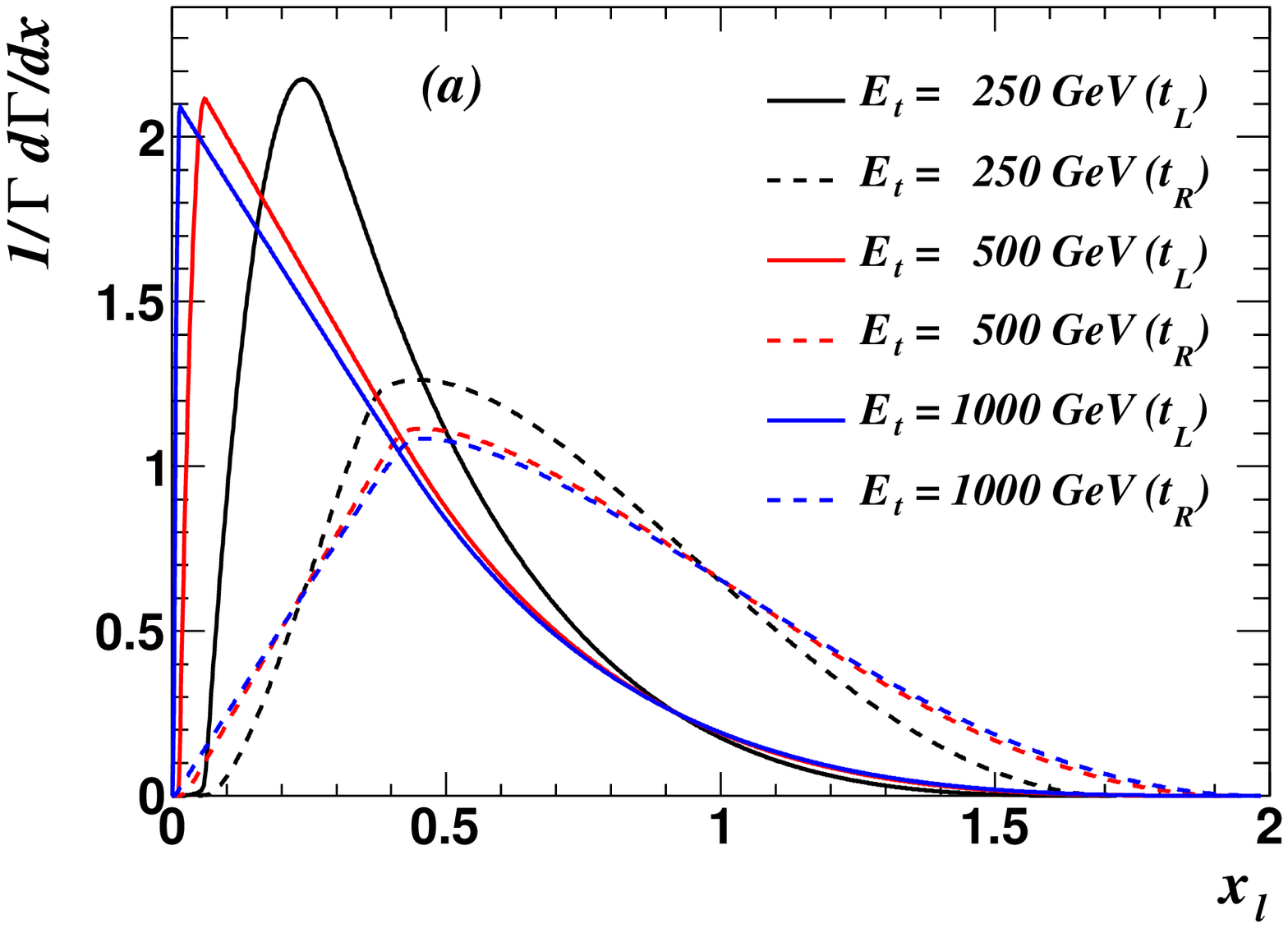}
\includegraphics[scale=0.4,clip]{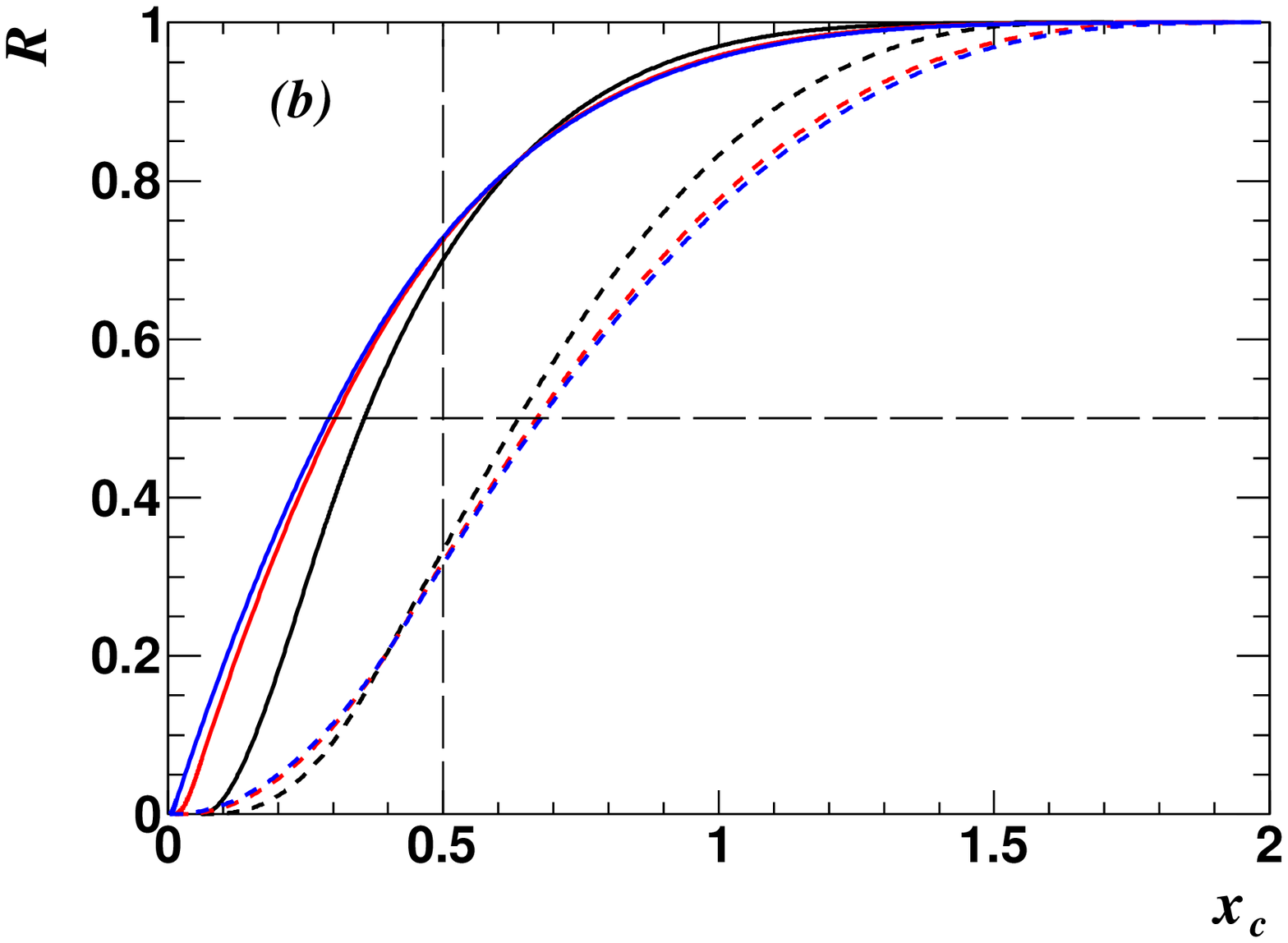}
\caption{
(a)Distributions of the energy fraction $x_{\ell}$ of a charged lepton from top quark decay for $E_{t}=250,~500,~1000~{\rm GeV}$;  (b) The ratio 
$\mathcal R$ as a function of the cut threshold $x_c$ for $E_{t}=250,~500,~1000~{\rm GeV}$. 
The solid lines represent left-handed top quark decay while the dashed lines represent right-handed top quarks.  
\label{fig:en_lep} }
\end{figure}
For a boosted top-quark with energy $E_t$, the distribution in the energy fraction $x_{\ell} \equiv 2 E_\ell / E_t$ of the charged lepton becomes   
\bea
&&\frac{d\Gamma(\hat{s}_t )}{dx}=\frac{\alpha_{W}^{2}m_{t}}{64\pi AB}\int_{\zmin}^{\zmax}
x\gamma^{2}[1-x\gamma^{2}(1-z\beta)]
\nonumber \\
&& \qquad
\left(1+\hat{s}_{t}\frac{z-\beta}{1-z\beta}\right)
\text{Arctan}\biggl[\frac{Ax\gamma^{2}(1-z\beta)}{B-x\gamma^{2}(1-z\beta)}\biggr]dz. ~~~~
\label{eq:energy}
\eea
Here, $A=\Gamma_{W}/m_{W}$ is the ratio of the $W$-boson width and $W$-boson mass, $B=m_{W}^{2}/m_{t}^{2}\approx 0.216$ is the ratio of the $W$-boson mass and top quark mass, and the limits of integration are $\zmin = {\rm max}[(1-1/\gamma^2 x)/\beta,-1]$, $\zmax = {\rm min}[(1-B/\gamma^2 x)/\beta,1]$ with $\gamma = E_t/m_t$ and $\beta = \sqrt{1-1/\gamma^{2}}$. 
The function $\text{Arctan}$ is defined as 
$\arctan(x)$ for $x\ge 0$ while $\pi+\arctan(x)$ for $x<0$.

Figure~\ref{fig:en_lep}(a) displays the normalized energy fraction of the charged lepton from Eq.~\ref{eq:energy} for three top quark energies, $E_t= (250~,500,~1000)~\rm{GeV}$, for both left-handed and right-handed top quark decay.  The important 
point~\cite{Czarnecki:1990pe,Schmidt:1992et} is that  
right-handed top-quarks $t_R$ (dashed curves) produce more energetic leptons than left-handed top-quarks $t_L$ (solid curves), 
with the difference becoming more pronounced with increasing $E_t$.  

We exploit the different dependence of $t_L$ and $t_R$ on $x_{\ell}$ shown in Fig~\ref{fig:en_lep}(a) to measure the top quark 
polarization~\cite{Zhang:2010kr}.     
We introduce a ratio $\mathcal{R}$ as a quantitative measure of the energy fractions of $t_L$ and $t_R$, 
\be
\mathcal{R}(x_c)=\frac{\displaystyle 1}{\displaystyle \Gamma} \int_{0}^{x_c}\frac{d\Gamma}{d x_\ell} dx_\ell
\equiv \frac{\Gamma(x_\ell<x_c)}{\Gamma}.  
\ee
This ratio is a function of the the cut threshold $x_c$ of the energy fraction $x_\ell$. 
We plot the  $\mathcal{R}(x_c)$ distribution in Fig.~\ref{fig:en_lep}(b) for three top quark energies, $E_t= (250~,500,~1000)~\rm{GeV}$, for both $t_L$ (solid curves) and $t_R$ (dashed curves).  
While the energy distribution Fig.~\ref{fig:en_lep}(a) varies with the top-quark energy, the $\mathcal{R}(x_c)$ distribution in 
Fig.~\ref{fig:en_lep}(b) shows much less dependence.   

An analytic expression can be derived for $\mathcal{R}(x_c)$ in the limit $\beta \to 1$.  It takes the form 
\be
\mathcal{R}(x_c)=\frac{3x_c(1-\lambda_t)}
{2(1+2B)}-\frac{3\lambda_t x_c^2(1-B+\ln B)}{2(1+2B)(1-B)^2},  
\label{eq:anal1}
\ee
for $x_c \in (0,~2B)$, and 
\bea
&&\mathcal{R}(x_c)=\frac{B^2(2B-3)}{(1+2B)(1-B)^2}+\frac{3x_c(1-\lambda_t)}{2(1-B)^2(1+2B)} \nonumber\\
&&-\frac{3x_c^2[1+2\lambda_t\ln(x_c/2)]}{4(1-B)^2(1+2B)}+\frac{x_c^3(1+3\lambda_t)}{8(1-B)^2(1+2B)}
\label{eq:anal2}
\eea
for $x_c\in (2B,~2)$, where $\lambda_t=(-1, +1)$ for $(t_L, t_R)$, respectively.  For small $x_c$, these expressions show that $\mathcal{R}(x_c)$ grows linearly with $x_c$ for $t_L$, whereas $\mathcal{R}(x_c)$ grows as  $x_c^2$ for $t_R$ ($\lambda_t =1$).

The analytic expressions Eqs.~\ref{eq:anal1} and~\ref{eq:anal2} also explain why the curves for $E_t=500~{\rm GeV}$ ($\beta=0.94$) and $E_t=1000~{\rm GeV}$ ($\beta=0.99$) almost overlap.  For an energetic top-quark, an important consequence is that the difference between $\mathcal{R}(x_c)$ for $t_L$ and $t_R$ is not sensitive to $E_t$, i. e., the mass splitting between the parent heavy resonance and the DM candidate, as long as the mass splitting is not too small.  The $t_L$ and $t_R$ curves are insensitive to the origin of the top quark in the collision, whether from a heavy fermion decay or from a scalar decay.  In other words, $\mathcal{R}(x_c)$ quantifies the top quark polarization but not the top quark origin.   Moreover, in order to extract NP signal events from SM backgrounds, one must normally impose a set of hard kinematic cuts on the leptons and jets in the final state.   These hard cuts force the top quark to be very energetic and thus to satisfy the limit $\beta \to 1$.  Therefore, another virtue of the $\mathcal{R}(x_c)$ variable is that the difference between 
$t_L$ and $t_R$ curves, do not vary with the hard cuts.

The ratio $\mathcal{R}(x_c)$ appears to show great promise for distinguishing $t_L$ and $t_R$.  However, even if it is insensitive to 
$E_t$, it presupposes reconstruction of the kinematics of the top quark (i.e., knowledge of $E_t$).   Moreover, until this point, we have not included the influence of the production dynamics of the top quark, including matrix elements and the convolution with parton distribution functions.  To prove our method useful, we must show that there are good estimators that can replace $E_t$.   To this end, we turn to an explicit calculation of top squark ($\tilde{t}$) pair production, $pp \to \tilde{t} \overline{\tilde{t}} X \to t \bar{t} \tilde{\chi} \tilde{\chi} X$.

\noindent{\bf Collider simulation:}~We perform a parton-level Monte Carlo simulation of top squark ($\tilde{t}$) pair production 
$pp \to \tilde{t} \overline{\tilde{t}} X \to t \bar{t} \tilde{\chi} \tilde{\chi} X$ to demonstrate that $\mathcal{R}$ remains useful for distinguishing $t_L$ and $t_R$ even when $E_t$ cannot be measured directly.   We assume the colored scalar $\tilde{t}$ decays entirely into 
$t\tilde{\chi}$ through the effective coupling
\be
\mathcal{L}_{\tilde{t}t\tilde{\chi}}= g_{\rm eff} \tilde{t} \tilde{\chi}(\cos\theta_{\rm eff} P_L + \sin\theta_{\rm eff} P_R) t, 
\label{eq:lagrange}
\ee
where the angle $\theta_{\rm eff}$ depends on the mass matrix mixings of the top squark and the neutralino sectors, and 
$P_{\rm L/R}$ is the usual left/right-handed projector.   
Our benchmark point has $m_{\tilde{t}}=360~{\rm GeV}$ and a representative DM mass $m_{\tilde{\chi}}=50~{\rm GeV}$.   
We simulate $\tilde{t}$ pair production with decay to a top quark pair plus dark matter candidates at the LHC with 8 TeV energy.   We demand that the top quark decays semi-leptonically, and that the anti-top quark decays hadronically, $\bar {t} \rightarrow 3 \rm{jets}$.   
The final state contains a lepton plus jets and large missing transverse energy $\met$.  
Two irreducible SM backgrounds, $t\bar{t}$ and $t\bar{t}Z$ production, are considered.
Both the signal and background processes are generated at leading order in MadGraph/MadEvent~\cite{Alwall:2007st} with CTEQ6L1 parton distribution functions~\cite{Pumplin:2002vw}.   The renormalization and factorization scales are chosen as $m_{\tilde{t}}$.  Momentum 
smearing effects are included through a Gaussian-type energy resolution.  We apply a set of basic acceptance cuts for the jets and  single lepton in the final state:  
$p_T(\ell)>20~{\rm GeV}$, $p_T(j)>25~{\rm GeV}$, $\left|\eta_{\ell,j}\right|<2.5$, $\Delta R_{jj,\ell j} > 0.4$, $\met > 25~{\rm GeV}$.
To suppress SM backgrounds, we impose a set of much harder cuts:  
$p_T(j_{\rm 1st})>50~{\rm GeV}$, $p_T(j_{\rm 2nd})>40~{\rm GeV}$, $\met > 100~{\rm GeV}$, $H_T>500~{\rm GeV}$, 
where $H_T$ is the scalar sum of the transverse energies of all objects in the event.  
After the hard cuts, the cut efficiency for the signal is about $44\%$  compared to the rate after the basic cuts. 
The $t\bar{t}$ background still dominates after the hard cuts,
and the $t\bar{t}Z$ background is negligible. 
In order to further suppress the SM background, we use the fact that $\met $ originates from the neutralino and neutrino in the signal events while from only the neutrino in the $t\bar{t}$ background. Hence, the neutrino longitudinal momentum $p_{\nu L} $ obtained from the 
$W$-boson on-shell condition  $m_{l \nu}^2 = m_W^2$, 
\begin{eqnarray}
p_{\nu L}  ={1 \over {2 p_{e T}^2}}
 \left( {A\, p_{e L} \pm E_e \sqrt{A^2  - 4\,{p}^{\,2}_{e T}\not{\!\!{\rm E}}_{T}^2}} \right),
\end{eqnarray}
is unphysical more often in the signal than in the background~\cite{Han:2008gy}.  Here 
$A = m_W^2 + 2 \,\vec{p}_{e T} \cdot \,\vec{\not{\!\!{\rm E}}_{T}} $. We then demand $A^2 - 4  {p}^{2}_{e T} \not{\!\!{\rm E}}_{T}^{2} \le 0$. 
We also impose a cut on the transverse mass of the charged lepton and missing energy,
$M_T = \sqrt{ 2 p_T^\ell \met (1 - \cos\phi)} \geq 100~{\rm GeV}$,
where $p_T$ is the lepton transverse momentum and $\phi$ is the angle in the transverse plane between $\vec{p}_T$ and $\vec{\not{\!\!{\rm E}}_{T}}$. Only about 0.00556\% of the $t\bar{t}$ events remain after all the cuts. The cross sections for the signal and main backgrounds are shown in Table \ref{cuttab} after branching fractions are included.   Using these cross sections, we find that the numbers of signal and background 
events are 130 and 22 at 8 TeV and 20~fb$^{-1}$ integrated luminosity, for a signal significance of $S/\sqrt B = 28$.  
\begin{table}
	\caption{Cross sections (in fb) for the signal and backgrounds processes at different cut levels, 
	including the decay branching fractions to the specific final states of interest.}	\label{cuttab}
\begin{tabular}{|c|c|c|c|c|c|c}
\hline\hline
 &  {\it Basic} & {\it $t_{had}$ recon.} & {\it Hard} & {\it $\met$ sol.} &  $\epsilon_{\rm cut}$  \tabularnewline 
\hline\hline
signal       & 22.26  & 18.46  & 8.87 & 6.51 &  11.6 \% \tabularnewline
 $t\bar{t}$   & 4347.08   & 3596.75 & 154.47 & 0.91 &  0.00556\%  \tabularnewline
 $t\bar{t}Z$ & 1.25   & 1.03  & 0.34 & 0.22 &  5.9 \%    \tabularnewline
\hline
\hline
\end{tabular}
\end{table}

In $\tilde{t}$ pair production the decay chains of $\tilde{t} \to t \tilde{\chi}$ and  $\tilde{\bar{t}} \to \bar{t} \tilde{\chi}$ have similar 
kinematics because the heavy $\tilde{t}$'s are not highly boosted.
In this work we investigate the energy of the anti top-quark as an estimator of the top quark energy, with the anti-top quark 
required to decay into three jets~\footnote{Another useful variable in the literature is $E_\ell/(E_\ell + E_b)$ where the $b$-jet and $\ell^+$ originate from the same top-quark decay~\cite{Shelton:2008nq}. }.  
We define a new energy fraction variable $x^\prime_{\ell}$,
\be
x^\prime_\ell = 2 E_{\ell} / E_{\bar{t}}.  
\ee
After convolution with the production cross section, a ratio $\mathcal{R}^\prime$ can be defined as 
\be
\mathcal{R}^\prime(x_c)=\frac{1}{\sigma({\rm tot})}{\displaystyle \int_{0}^{x_c}\frac{d\sigma}{d x^\prime_\ell} dx^\prime_\ell} 
\equiv \frac{\sigma(x^\prime_\ell<x_c)}{\sigma({\rm tot})},
\ee
where $d\sigma/d x_\ell^\prime$ is the differential cross section, and $x^\prime_c$ is the cut threshold of the energy fraction $x^\prime_\ell$. 

We use a $\chi^2$-template method based on the $W$ boson and 
top quark masses to select the three jets from the hadronic decay of the anti-top quark.  
For each event we pick the combination which minimizes the following $\chi^2$:
\be
\chi^2 = \frac{(m_W-m_{jj})^2}{\Delta m_W^2}  + \frac{(m_{t}-m_{jjj})^2}{\Delta m_t^2},
\ee
where $\Delta m_W$ and $\Delta m_t$ are the width of the $W$-boson and the top quark, respectively.
The efficiency of this method is 84\%.
\begin{figure}
\includegraphics[scale=0.4,clip]{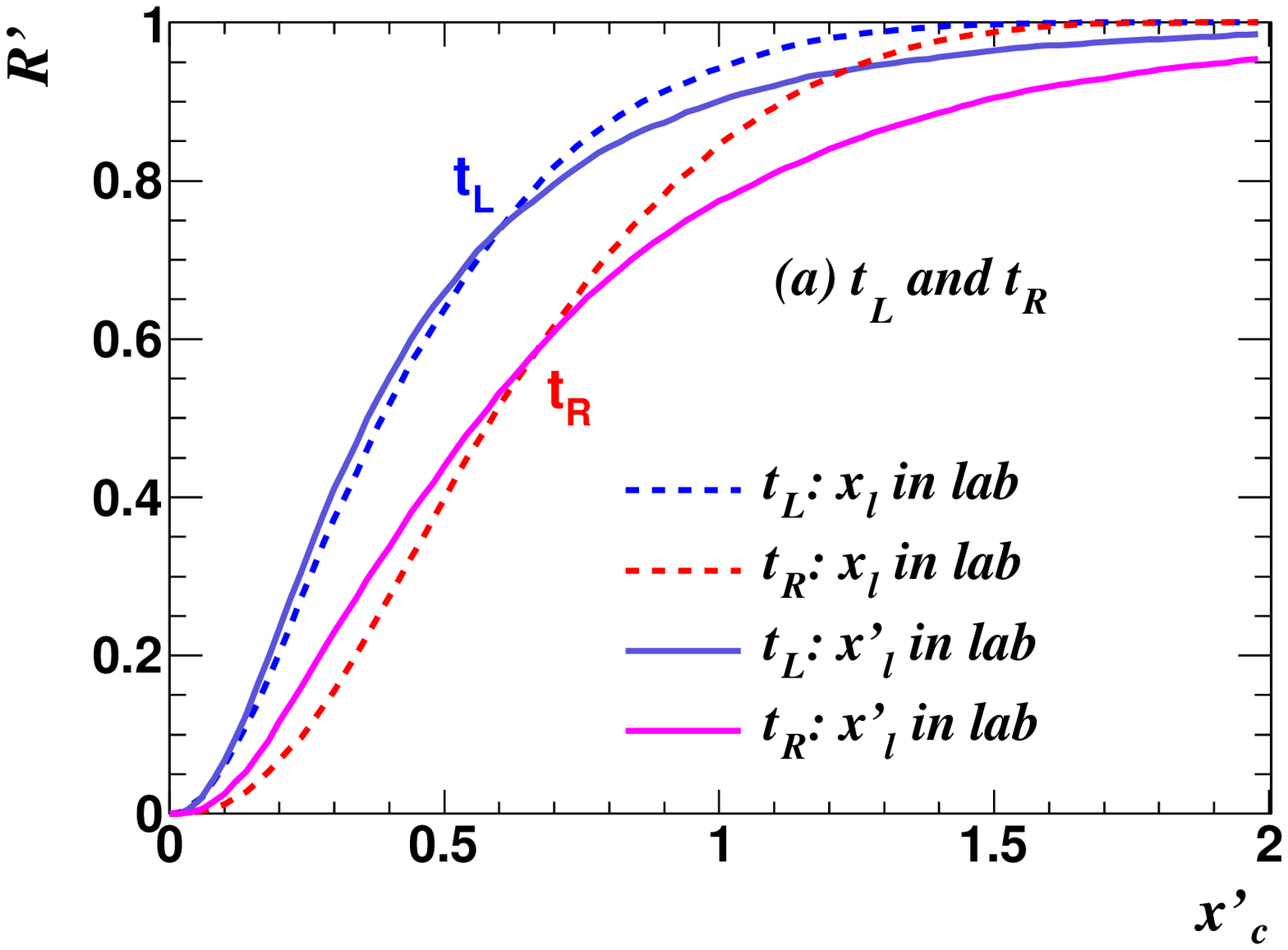}
\includegraphics[scale=0.4,clip]{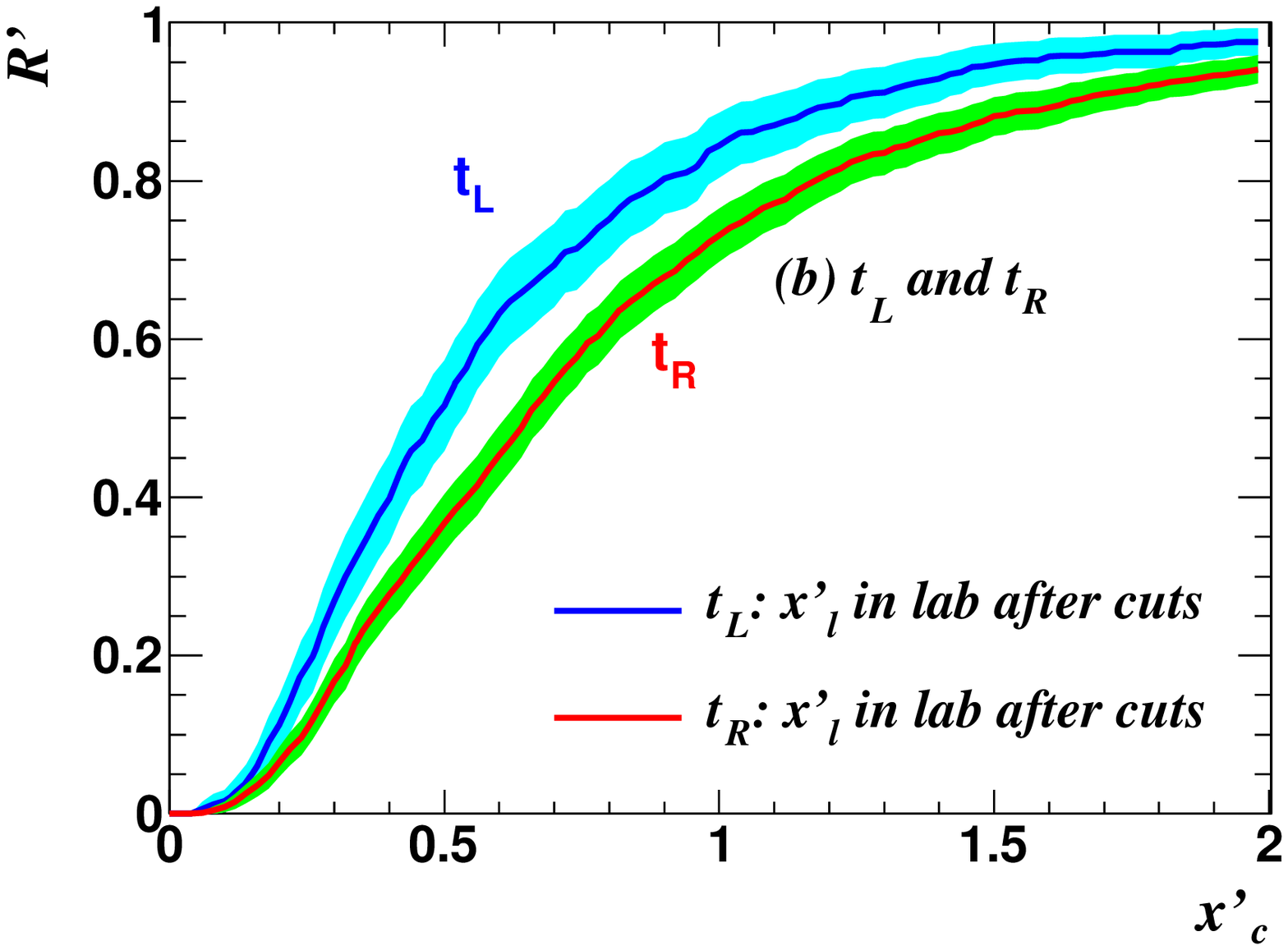}
\caption{(a) The $\mathcal R$ distributions as a function of the cut threshold $x^\prime_c$ for a 350~GeV $\tilde{t}$ quark with 
pure right-handed or left-handed couplings  
at the LHC with 8 TeV energy. The lepton energy fraction is evaluated in the lab frame from either $\bar{t}_{\rm had}^{\rm rec}$ 
($x'_{\ell}$) or the top quark energy $t_{\rm lep}$ ($x_{\ell}$).   
(b) The statistical uncertainty band of $\mathcal R$ is shown for both $t_L$ and $t_R$ at the $1\sigma$ confidence level 
for an assumed $20$ fb$^{-1}$ of integrated luminosity.   
\label{fig:mssm_cut} }
\end{figure}
After the antitop quark energy is reconstructed in the lab frame, $\mathcal R^\prime$ can be obtained with its cut threshold $x^\prime_c$ dependence.

Armed with both the Monte Carlo level momenta and the reconstructed momenta, we perform several comparisons to evaluate how faithful the $\mathcal R^\prime$ distribution is to the true $\mathcal R$.   At the Monte Carlo level, $ {t}_{\rm lep}$ and $\bar{t}_{\rm had}$ are known in the center-of-mass (cms) and lab frames.   Our comparisons show that $\mathcal R$ defined with $ {t}_{\rm lep}$ is not sensitive to the boost from the cms to laboratory frame, whereas $\mathcal R$ defined by $\bar{t}_{\rm had}$ shows a slight dependence.  
We compute the ratio $\mathcal{R}$ defined from the energy of the $ {t}_{\rm lep}$ and $\bar{t}_{\rm had}$.   At the detector simulation 
level, only the four-momentum of $\bar{t}_{\rm had}$ can be reconstructed, denoted $\bar{t}_{\rm had}^{\rm rec}$.  Some of our results are  
compared in Fig.~\ref{fig:mssm_cut} (a) for choices $\sin\theta_{\rm eff}=1$ and $\cos\theta_{\rm eff}=1$ in Eq.~\ref{eq:lagrange}.   With 
$\sin\theta_{\rm eff}=1$ ($\cos\theta_{\rm eff}=1$) the top quark is mainly right-handed (left-handed), and we label the curves by $t_R$ 
($t_L$).   There is some difference between the $\mathcal R$ distributions for $ {t}_{\rm lep}$ and $\bar{t}_{\rm had}$, but the essential 
features are preserved.  We conclude that $x_c^\prime$ is a good variable when $x_c$ cannot be obtained.   We also investigate the cut dependence of $\bar{t}^{\rm rec}_{\rm had}$ at the reconstruction level, whether basic or hard, and find that $\mathcal{R}$ is not sensitive to the cuts; the curves for the loose cuts and the hard cuts overlap.  Lastly, comparing $\mathcal{R}$ at the Monte Carlo level and at the reconstruction level, we see a slight downward shift for both $t_L$ and 
$t_R$.   This effect arises because the $p_T$ cuts on the lepton reduce the number of events with $x^\prime_{\ell} < x^\prime_c$.

The results in Fig.~\ref{fig:mssm_cut} (a) establish that $x_c^\prime$ is a suitable variable and that $\mathcal R^\prime$ serves as a good substitute for $\mathcal R$.    To show that the difference between the expectations for $t_L$ and $t_R$ can be observed, we include statistical uncertainties on the $\mathcal R$ variable.   Expressing  
\be
\mathcal{R}^\prime(x^\prime_c)= \frac{N(x^\prime_\ell < x_c^\prime)}{N({\rm tot})},
\ee
where $N({\rm tot})$ is the total event number after cuts, and $N(x^\prime_\ell < x_c^\prime)$ ($N(x^\prime_\ell > x_c^\prime)$) represents events with $x^\prime_\ell < x_c^\prime$ ($x^\prime_\ell > x_c^\prime$), we derive the 
standard statistical uncertainty 
\bea
\delta\sigma (x^\prime_c) & = &  \frac{1}{N({\rm tot})}\sqrt{\frac{N(x^\prime_\ell < x_c^\prime) N(x^\prime_\ell > x_c^\prime)}{N({\rm tot})}}.
\eea
The statistical uncertainties of the $\mathcal R$ distributions for $t_L$ and $t_R$ are shown in Fig.~\ref{fig:mssm_cut} (b) at the $1\sigma$ confidence level with $20$ fb$^{-1}$ integrated luminosity.  The distinction is evident between $L$ and $R$.  The $t_L$ 
curve increases much faster than the $t_R$ curve.   The pattern of the distributions of $\mathcal{R}$ can be used to identify the top quark polarization. In order to distinguish $t_L$ from an unpolarized top-quark, we define 10 bins in $x_c^{\prime}$ within the range $(0.3, 1.3)$ and calculate 
$\chi^2$ per degree-of-freedom (d.o.f) for the difference $\mathcal{R}^\prime_L (x_c^\prime)-\mathcal{R}^\prime_{0}(x_c^\prime)$: 
\be
\chi^2/{\rm d.o.f} = \frac{1}{10} \sum_{i=1}^{10} \left(\frac{\mathcal{R}^\prime_L (x_c^{\prime i})-\mathcal{R}^\prime_{0}(x_c^{\prime i})}{\delta \sigma_L(x_c^i)}\right)^2 .
\ee
The subscript $``0"$ denotes an unpolarized top-quark.  The result for an unpolarized top-quark ($t_0$) is 
\be
\mathcal{R}^\prime_0 (x_c^\prime)= \left(\mathcal{R}^\prime_L(x_c^\prime) + \mathcal{R}^\prime(x_c^\prime)\right)/2~.
\ee  
After all the cuts, about 87 (101) 
 signal events are needed to distinguish $t_L$ ($t_R$) from $t_0$ at 95\% confidence level (C.L.) in the absence of background, with only about 23 signal events to discriminate $t_R$ from $t_L$.

Thus far in this section, we assume the $t$-$\tilde{t}$-$\tilde{\chi}$ coupling is completely left-handed or right-handed, but in general the coupling is a mixture of both.  Once data are obtained, we could use the $\mathcal{R}^\prime$ curves shown in Fig.~\ref{fig:mssm_cut}(b) 
as templates in fits to these data to extract $\theta_{\rm eff}$ and shed light on the nature of top squark mixing.

\noindent{\bf Other implications:}~Our method can be applied to several NP models.  We performed a detailed simulation of pair production of a $T$-odd top quark partner ($T_-$) in
the Littlest Higgs Model with T-parity (LHT), $pp\to T_-\overline{T}_- X \to t\bar{t} A_H A_H X$,  
where $A_H$ is the $T$-odd photon partner.     
Our numerical results are very similar to those shown for $t_R$ in Fig.~\ref{fig:mssm_cut}.    
Verification of mainly right-handed polarization would provide a powerful check of the model~\cite{Cao:2006wk}.  Another example is the leptophobic $Z^\prime$ boson, which couples only to the SM quarks. 
The top quark polarization could be used to probe the handedness of the $Z^\prime$-$q$-$q$ coupling which is sensitive to how the SM quarks are gauged under the new gauge symmetry~\cite{Berger:2011hn}.

\noindent{\bf Acknowledgments}~The work by E.L.B. and H.Z. is supported in part by the U.S.
DOE under Grant No.~DE-AC02-06CH11357. H.Z. is also supported by DOE under the Grant No.~DE-FG02-94ER40840. 
Q.H.C. is supported by the National Natural Science Foundation of China under Grant No. 11245003.
J.H.Y. is supported by the U.S. National Science Foundation 
under Grant No. PHY-0855561.


\begin{thebibliography}{199}
	
\bibitem{Kane:1991bg}
  G.~Mahlon and S.~J.~Parke,
  Phys.\ Rev.\ D {\bf 53}, 4886 (1996)
  [hep-ph/9512264];
  G.~L.~Kane, G.~A.~Ladinsky and C.~P.~Yuan,
  Phys.\ Rev.\ D {\bf 45}, 124 (1992).

\bibitem{Cao:2012rz} 
  J.~Cao, C.~Han, L.~Wu, J.~M.~Yang and Y.~Zhang,
  arXiv:1206.3865;
  Y.~Bai, H.~-C.~Cheng, J.~Gallicchio and J.~Gu,
  arXiv:1203.4813;
  X.~-J.~Bi, Q.~-S.~Yan and P.~-F.~Yin,
  Phys.\ Rev.\ D {\bf 85}, 035005 (2012);
  T.~Plehn, M.~Spannowsky and M.~Takeuchi,
  JHEP {\bf 1208}, 091 (2012)
  [arXiv:1205.2696 [hep-ph]];
  D.~S.~M.~Alves, M.~R.~Buckley, P.~J.~Fox, J.~D.~Lykken and C.~-T.~Yu,
  arXiv:1205.5805 [hep-ph].


\bibitem{Perelstein:2008zt} 
  M.~Perelstein and A.~Weiler,
  JHEP {\bf 0903}, 141 (2009).


\bibitem{Czarnecki:1990pe} 
  A.~Czarnecki, M.~Jezabek and J.~H.~Kuhn,
  Nucl.\ Phys.\ B {\bf 351}, 70 (1991).


\bibitem{Schmidt:1992et} 
  C.~R.~Schmidt and M.~E.~Peskin,
  Phys.\ Rev.\ Lett.\  {\bf 69}, 410 (1992).
    
\bibitem{Zhang:2010kr} 
  H.~Zhang, E.~L.~Berger, Q.~-H.~Cao, C.~-R.~Chen and G.~Shaughnessy,
  Phys.\ Lett.\ B {\bf 696}, 68 (2011);
  E.~L.~Berger, Q.~-H.~Cao, C.~-R.~Chen, G.~Shaughnessy and H.~Zhang,
  Phys.\ Rev.\ Lett.\  {\bf 105}, 181802 (2010).
  
  
\bibitem{Alwall:2007st} 
  J.~Alwall, P.~Demin, S.~de Visscher, R.~Frederix, M.~Herquet, F.~Maltoni, T.~Plehn and D.~L.~Rainwater {\it et al.},
  JHEP {\bf 0709}, 028 (2007).

\bibitem{Pumplin:2002vw} 
  J.~Pumplin, D.~R.~Stump, J.~Huston, H.~L.~Lai, P.~M.~Nadolsky and W.~K.~Tung,
  JHEP {\bf 0207}, 012 (2002).
          


\bibitem{Han:2008gy} 
  T.~Han, R.~Mahbubani, D.~G.~E.~Walker and L.~-T.~Wang,
  JHEP {\bf 0905}, 117 (2009).

\bibitem{Shelton:2008nq} 
  J.~Shelton,
  Phys.\ Rev.\ D {\bf 79}, 014032 (2009).

\bibitem{Cao:2006wk} 
  Q.~-H.~Cao, C.~S.~Li and C.~-P.~Yuan,
  Phys.\ Lett.\ B {\bf 668}, 24 (2008);
  M.~M.~Nojiri and M.~Takeuchi,
  JHEP {\bf 0810}, 025 (2008).

    
\bibitem{Berger:2011hn} 
  E.~L.~Berger, Q.~-H.~Cao, C.~-R.~Chen and H.~Zhang,
  Phys.\ Rev.\ D {\bf 83}, 114026 (2011). 
  

  
\end{thebibliography}
\end{document}